\documentclass[preprint,12pt]{elsarticle}
\usepackage{graphics}
\usepackage{mathrsfs}
\usepackage{amsmath}

\def\Vec#1{\mbox{\boldmath $#1$}}
\journal{Physica A}

\begin{document}

\begin{frontmatter}

\title{A new perspective to formulate \\ a dissipative thermo field dynamics}


\author[1]{Yoichiro Hashizume}
\address[1]{Tokyo University of Science,\\ 6-3-1 Niijuku, Katsushika-ku, Tokyo, 125-8585, Japan}
\ead{hashizume@rs.tus.ac.jp}
\author[2]{Masuo Suzuki}
\address[2]{Computational Astrophysics Laboratory, RIKEN,\\ 2-1 Hirosawa, Wako, Saitama, 351-0198, Japan}
\ead{masuo.suzuki@riken.jp}
\author[1]{Soichiro Okamura}
\ead{sokamura@rs.kagu.tus.ac.jp}

\begin{abstract}
In the present study, we propose a new perspective on thermal dissipation based on the thermo field dynamics.
From the view point of the renormalization theory, there appear effective interactions between the original and tilde spaces on thermo field dynamics with reducing thermal disturbances.
This study yields the equivalence of the following two pictures, namely such a spin system with a random field due to a heat bath as is defined in a Hilbert space and a finite-size system with effective interactions defined in a double Hilbert space. 
The correspondence of the above two systems yields such perspective that the thermal disturbance is described by the effective non-Hermitian interactions.
\end{abstract}


\end{frontmatter}

\section{Introduction}
\label{intro}

Recently, thermo field dynamics (TFD) [1-4] played a very useful and important role to study entanglements [5-11], Kondo effects \cite{12}, the entropy effect of a black hole \cite{13,14} and non-equilibrium phenomena [5,6,15-17].
We have to introduce a tilde space in order to study statistical properties using TFD.
Actually, the state vector $|\Psi\rangle$, namely ``TFD state vector", is defined [1-4] as 
\begin{equation}
|\Psi\rangle = \rho^{1/2} |I\rangle {\text{ and }} |I\rangle =\sum_{n}|n\rangle\otimes|\tilde{n}\rangle\equiv \sum_{n}|n, \tilde{n}\rangle \label{eq1} 
\end{equation}
with the density matrix $\rho$ of the relevant system.
The bases $|n\rangle$ (or $|\tilde{n}\rangle$) are the eigenstates of the Hamiltonian (or $\tilde{H}$ in the tilde space) with the eigenvalues $\{E_n\}$, namely $\mathcal{H}|n\rangle =E_n|n\rangle$ (or $\tilde{\mathcal{H}}|\tilde{n}\rangle =E_n|\tilde{n}\rangle$). 
One of the present authors \cite{18,19} proved that the state $|I\rangle$ itself is invariant for any representation, namely
\begin{equation}
|I\rangle =\sum_{\alpha}|\alpha\rangle\otimes|\tilde{\alpha}\rangle\equiv \sum_{\alpha}|\alpha, \tilde{\alpha}\rangle \label{eq1-1} 
\end{equation}
where $\{\alpha\}$ is an arbitrary orthogonal complete set.
Then we can easily derive \cite{4,18,19} the thermal average as
\begin{equation}
\langle \Psi | Q | \Psi\rangle =\sum_{\alpha,\alpha'} \langle \alpha, \tilde{\alpha}| \rho^{1/2} Q \rho^{1/2} | \alpha', \tilde{\alpha}'\rangle={\text{Tr}}\rho Q=\langle Q \rangle \label{eq2} 
\end{equation}
for the physical quantity $Q$.
Especially, once the tilde vector $|\tilde{\alpha} \rangle$ satisfies the relation $|\tilde{\alpha}'\rangle=\sum_{\alpha}U_{\alpha',\alpha}^{*}|\tilde{\alpha}\rangle$ for $|\alpha' \rangle=\sum_{\alpha}U_{\alpha',\alpha} |\alpha\rangle$, one can find clearly the ``general representation theorem" \cite{4,18,19}.
The general representation theorem means not only the thermal average $\langle Q \rangle$ but also the representations of TFD state vectors are independent of the bases $\{|\alpha \rangle \}$, because \cite{4,18,19}
\begin{equation}
|I'\rangle =\sum_{\alpha'}|\alpha', \tilde{\alpha}'\rangle=\sum_{\alpha'}\sum_{\alpha_1,\alpha_2}U_{\alpha',\alpha_1}^{*}U_{\alpha',\alpha_2}|\alpha_1, \tilde{\alpha}_2\rangle=\sum_{\alpha_1,\alpha_2} \delta_{\alpha_1,\alpha_2}|\alpha_1, \tilde{\alpha}_2\rangle =|I\rangle. \label{eq3}
\end{equation}
Furthermore, this general representation theorem derives \cite{4,18,19} the time development equation
\begin{equation}
i\hbar\frac{\partial }{\partial t} |\Psi(t)\rangle = \hat{\mathcal{H}}|\Psi(t)\rangle \label{eq4}
\end{equation}
of the TFD state vector $|\Psi(t)\rangle$ using the von Neumann equation
\begin{equation}
i\hbar\frac{\partial }{\partial t}\rho(t)=[\mathcal{H},\rho(t)]. \label{eq5}
\end{equation}
This is one of the important merits of the general representation theorem \cite{4,18,19}.
Here $\hat{\mathcal{H}}$ is defined as $\hat{\mathcal{H}}=\mathcal{H}-\tilde{\mathcal{H}}$.
Additionally, we have introduced \cite{6} the extended density matrix $\hat{\rho}\equiv |\Psi\rangle\langle \Psi|$ and found a simple way to study the entanglement directly even in a dissipative system.  
On the above discussions of TFD, a Hilbert space defined by a set of bases $\{ |\alpha\rangle \}$ (namely ``original space") is extended to a double Hilbert space which is defined by a direct product space $\{ |\alpha\rangle\otimes |\tilde{\alpha} \rangle \}$ \cite{4}.
Note that the tilde space $\{ |\tilde{\alpha} \rangle \}$ needs to be isomorphic to the original space $\{ |\alpha\rangle\}$.

In this way, the tilde space is introduced in a mathematical point of view, and then its physical meanings are not so clear.
Nonetheless, the double Hilbert space is very useful practically.
For example, it was applied to the density matrix renormalization group (DMRG) analysis and multi-scale entanglement renormalization ansatz (MERA) \cite{8,9} at finite temperatures \cite{7,10,11}, and it was used  to study Kondo effect \cite{12} and to study a black hole entropy \cite{13,14}.
There are some common features in these previous studies, in which the thermal effect can be treated by introducing the tilde space (or double Hilbert space).
In the present study, we try here to clarify how the thermal noises are renormalized into TFD pictures.
This study does not clarify the origin of thermal disturbances from the viewpoint of first principles but clarifies phenomenologically the renormalization scheme of thermal noises using TFD.

In the following section, by treating a simple example, we obtain an effective interaction between the original and tilde spaces.
It may be trivial that effective interactions appear owing to renormalization procedure.
However, these effective interactions enable us to understand a plausible treatment of a heat bath and they support the previous suggestions [15,20-22] for the dissipative formulation of TFD.
In section 3, the present new perspective is applied to many-body systems.
Summary and discussions are included in section 4.

\section{Observation of new perspective on a simple example with a thermal noise}

In this section, we present a new perspective of tilde space using a simple example.
In 1961, Schwinger \cite{23} showed that effects of a heat bath can be represented by a thermal noise, using a fluctuation propagator of quantum oscillators.
Thus, we adopt a single spin system with the thermal noise $R(t)$ described by the following Hamiltonian including a spin variable $S_0$;
\begin{equation}
\mathcal{H}=\mathcal{H}_0-JR(t)S_0, \label{eq6}
\end{equation}
where the thermal noise $R(t)$ is regarded as the white Gaussian noise, namely
\begin{equation}
\langle R(t) \rangle_R=0 {\text{ and }} \langle R(t)R(t') \rangle_R =\epsilon \delta(t-t'). \label{eq7}
\end{equation}
$\mathcal{H}_0$ is a Hamiltonian of interest without noise. Here the notation $\langle A \rangle_R$ denotes the random average of the stochastic parameter $A$, and the noise intensity is denoted by the parameter $\epsilon$.
The constant real number $J$ denotes a kind of coupling constant.
For the Hamiltonian (\ref{eq6}), the tilde Hamiltonian $\tilde{\mathcal{H}}$ \cite{18} is obtained by
\begin{equation}
\tilde{\mathcal{H}}=\tilde{\mathcal{H}}_0-JR(t)\tilde{S}_0. \label{eq8}
\end{equation}
It is important that the same thermal noise $R(t)$ is used both in the tilde Hamiltonian (\ref{eq8}) and in the original Hamiltonian (\ref{eq6}), because the tilde space needs to be isomorphic to the original space.
Then the TFD Hamiltonian $\hat{\mathcal{H}}$ in the present double Hilbert space is defined by
\begin{equation}
\hat{\mathcal{H}}=\mathcal{H}-\tilde{\mathcal{H}}=\hat{\mathcal{H}}_0-JR(t)(S_0-\tilde{S}_0)\equiv \hat{\mathcal{H}}_0+\hat{\mathcal{H}}'(t).\label{eq9}
\end{equation}
The time dependence of the TFD state vector $|\Psi(t)\rangle$ is expressed as
\begin{equation}
i\hbar \frac{\partial}{\partial t}|\Psi(t)\rangle =\hat{\mathcal{H}}|\Psi(t)\rangle=(\hat{\mathcal{H}}_0+\hat{\mathcal{H}}'(t))|\Psi(t)\rangle \label{eq10}
\end{equation}
from Eq.(\ref{eq4}).
The formal solution of Eq.(\ref{eq10}) is given by
\begin{equation}
|\Psi(t)\rangle = \hat{U}(t)|\Psi(t=0)\rangle \label{eq11}
\end{equation}
using the operator $\hat{U}(t)$.
This time-development operator $\hat{U}(t)$ includes the thermal noise $R(t)$.
Clearly, by reducing the random operator $R(t)$, we can obtain such an effective interaction between $S_0$ and $\tilde{S}_0$ as includes thermal dissipation.
This yields, as will be seen below, the new point of view that introducing the effective interaction enables us to study dissipative systems.

The time development operator $\hat{U}(t)$ is represented in the following time ordered product;
\begin{equation}
\hat{U}(t)=\hat{U}_0(t)\left\{ 1+\sum_{n=1}^{\infty}\left( \frac{1}{i\hbar} \right)^n \int_{0}^{t}dt_{1}\int_{0}^{t_1}dt_{2}\cdots\int_{0}^{t_{n-1}}dt_{n} \hat{\mathcal{H}}'(t_1) \hat{\mathcal{H}}'(t_2)\cdots \hat{\mathcal{H}}'(t_{n})\right\}, \label{eq12}
\end{equation}
where $\hat{U}_0(t)\equiv \exp[-i\mathcal{H}_0t/\hbar]$ does not include the thermal noise $R(t)$.
Then, taking the random average $\langle \hat{\mathcal{H}}'(t_1) \hat{\mathcal{H}}'(t_2)\cdots \hat{\mathcal{H}}'(t_{n}) \rangle _{R}$, we obtain the reduced operator $\langle \hat{U}(t)\rangle_R$.
In the present study, higher order commutations are not so important for our aims to propose a simple picture of the tilde space as a heat bath.
This is because we apply a decoupling approximation (mean-field approximation) as
\begin{align}
&\langle \hat{\mathcal{H}}'(t_1) \hat{\mathcal{H}}'(t_2)\cdots \hat{\mathcal{H}}'(t_{n}) \rangle _{R} \notag
\\
&\simeq \langle \hat{\mathcal{H}}'(t_1) \hat{\mathcal{H}}'(t_2)\rangle _{R}\langle \hat{\mathcal{H}}'(t_3) \hat{\mathcal{H}}'(t_4)\rangle _{R}\cdots \langle \hat{\mathcal{H}}'(t_{2k-1}) \hat{\mathcal{H}}'(t_{2k})\rangle _{R};\quad 2k\equiv n \notag
\\
&=\left[ \epsilon J^2(S_0-\tilde{S}_0)^2 \right]^k \delta (t_{1}-t_{2})\cdots \delta (t_{2k-1}-t_{2k}).\label{eq13}
\end{align}
Using Eqs.(\ref{eq12}) and (\ref{eq13}), the reduced time development operator $\langle \hat{U}(t)\rangle_R$ is obtained as
\begin{equation}
\langle \hat{U}(t) \rangle_R = \hat{U}_0(t)\exp \left[ -\frac{\epsilon J^2}{2\hbar^2}(S_0-\tilde{S}_0)^2 t \right] = e^{-i \hat{\mathcal{H}}_0t/\hbar}e^{\gamma t S_0\tilde{S}_0-\gamma t}, \label{eq14}
\end{equation}
where the dissipative constant is denoted by $\gamma \equiv \epsilon J^2 /\hbar^2$.

As shown in Eq.(\ref{eq14}), there appears an effective interaction between $S_0$ and $\tilde{S}_0$ as $\gamma t S_0 \tilde{S}_0$.
This interaction depends on the parameter $\epsilon $ and $J$.
Consequently, this kind of effective interaction does appear only in the dissipative system with a heat bath.
Once thermal noises are reduced, the effective TFD Hamiltonian $\hat{\mathcal{H}}_{\text{eff}}$ does not include any thermal noise explicitly.
Then we can conclude that dissipative effects are renormalized into the effective interactions between the original and tilde spaces.
These contexts become much clear to consider the time differentials.
Using the notations
\begin{equation}
S_0(t)\equiv e^{-i \mathcal{H}_0 t/\hbar}S_0 e^{i \mathcal{H}_0 t/\hbar} {\text{ and }} \tilde{S}_0(t)\equiv e^{-i\tilde{\mathcal{H}}_0 t/\hbar}\tilde{S}_0 e^{i\tilde{\mathcal{H}}_0 t/\hbar}, \label{eq15}
\end{equation}
the time differential of the dissipative TFD state vector $|\Psi^{\text{diss}}(t) \rangle$ is derived as
\begin{align}
\frac{\partial}{\partial t}|\Psi^{\text{diss}}(t) \rangle &=\left(\frac{\partial}{\partial t}\langle \hat{U}(t)\rangle_{R} \right)|\Psi^{\text{diss}}_0\rangle\notag
\\
&=\left(\frac{1}{i\hbar}\hat{\mathcal{H}}_0+\hat{\Lambda}_0(t) \right)|\Psi^{\text{diss}}(t) \rangle \label{eq16}
\end{align}
for the initial condition $|\Psi(t=0)\rangle\equiv |\Psi_0\rangle=|\Psi^{\text{diss}}(t=0) \rangle=|\Psi^{\text{diss}}_0 \rangle$, where the effective interaction $i\hbar \hat{\Lambda}_0(t)$ is defined as
\begin{equation}
i\hbar \hat{\Lambda}_0(t)\equiv i\hbar \gamma \left( S_0(t)\tilde{S}_0(-t)-1 \right).\label{eq17}
\end{equation}
Eq.(\ref{eq16}) is nothing but the dissipative formulation proposed by one of the authors (M.S.) \cite{15}.
Now we clarify physical meanings of this dissipative term $\hat{\Lambda}_0(t)$.
As shown in Eq.(\ref{eq17}), this dissipative term $\hat{\Lambda}_0(t)$ includes an effective interaction which is denoted by a convolution form.
This convolution form comes from the definition of the TFD Hamiltonian $\hat{\mathcal{H}}\equiv \mathcal{H}-\tilde{\mathcal{H}}$ on the time development equation (\ref{eq10}) of TFD state vectors.
And then, this effective interaction is peculiar to dissipative TFD.
Of course, when there is no dissipation, the effective interaction $i\hbar \hat{\Lambda}_0(t)$ vanishes for $\epsilon \to 0$, and Eq.(\ref{eq16}) becomes the ordinal (non-dissipative) time developments as shown in Eq.(\ref{eq4}).
From Eqs.(\ref{eq16}) and (\ref{eq17}), the effective TFD Hamiltonian $\hat{\mathcal{H}}_{\text{eff}}$ is expressed as
\begin{equation}
\hat{\mathcal{H}}_{\text{eff}}=\hat{\mathcal{H}}_0+i\hbar \hat{\Lambda}_0(t). \label{eq18}
\end{equation}
Clearly, it is non-Hermitian.
This non-Hermitian framework is regulated \cite{20} by the physical constraints, assuming the initial and final conditions of relaxation properties and thermal state conditions.
In the present study, we have reinterpreted this framework by reducing the thermal disturbance applied to the thermal state.
This study yields the equivalence of the following two pictures, namely such a spin system with a random field due to a heat bath as is defined in a Hilbert space and a finite-size system with effective interactions defined in a double Hilbert space. 
The correspondence of these two systems yields such perspective that the thermal disturbance is described by the effective non-Hermitian interactions.
The alternative relation between the original and TFD spaces is clarified as shown in Fig.1.

\begin{figure}[hbpt]
\begin{center}
\includegraphics[width=12cm]{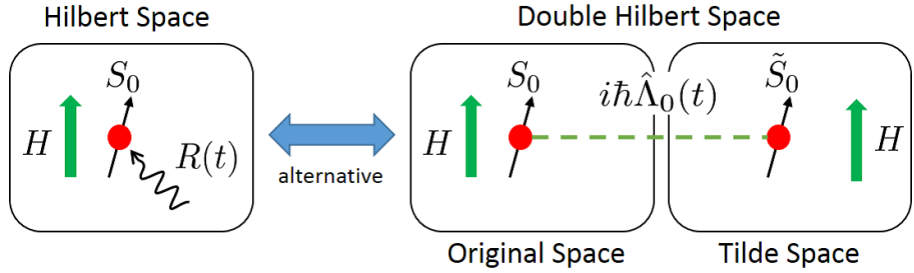}
\end{center}
\caption{Schematic picture of the correspondence between the dissipative system with a heat bath and the effectively interacting TFD system. In this figure, we assume $\mathcal{H}_0=-\mu_{\text{B}}\Vec{H}\cdot\Vec{S}_0$. The effective interaction $\hat{\Lambda}_0(t)$ appears with reducing the thermal disturbance $R(t)$.}
\label{fig1}
\end{figure}

Furthermore, from Eq.(\ref{eq16}), we can easily obtain the dissipative von Neumann equation
\begin{equation}
\frac{\partial}{\partial t}\hat{\rho}_{\text{diss}}(t)=\frac{1}{i\hbar}[\hat{\mathcal{H}}_0,\hat{\rho}_{\text{diss}}(t)]+\hat{\Lambda}_0(t)\hat{\rho}_{\text{diss}}(t)+\hat{\rho}_{\text{diss}}(t)\hat{\Lambda}_0(t),\label{eq19}
\end{equation}
using the extended density matrix \cite{6} $\hat{\rho}_{\text{diss}}(t)\equiv |\Psi^{\text{diss}}(t)\rangle\langle \Psi^{\text{diss}}(t)|$.
It also corresponds to the previous study \cite{15} by one of the authors (M.S.).

It may be interesting to describe explicitly the TFD state vector $|\Psi(t)\rangle$ for a simple example.
Here we assume the Hamiltonian $\mathcal{H}_0$ as
\begin{equation}
\mathcal{H}_0=-\mu_{\text{B}}HS_0^z.\label{eq20}
\end{equation}
The initial condition $|\Psi_0\rangle\equiv |\Psi(t=0)\rangle$ is assumed as
\begin{equation}
|\Psi_0\rangle=\alpha_{1}|+,\tilde{+}\rangle + \alpha_{2}|+,\tilde{-}\rangle + \alpha_{3}|-,\tilde{+}\rangle + \alpha_{4}|-,\tilde{-}\rangle \label{eq21}
\end{equation}
with the real numbers $\{\alpha_{1},\alpha_{2},\alpha_{3},\alpha_{4}| \sum_{j=1}^4 \alpha_{j}^2=1\}$.
Thus the dissipative TFD state vector $|\Psi^{\text{diss}}(t) \rangle$ is obtained as
\begin{align}
|\Psi^{\text{diss}}(t) \rangle &= \langle \hat{U}(t)\rangle_{R} |\Psi_0\rangle \notag
\\
&= (\alpha_{1}|+,\tilde{+}\rangle + \alpha_{4}|-,\tilde{-}\rangle)+e^{-2\gamma t}(e^{-2i\omega t}\alpha_{2}|+,\tilde{-}\rangle + e^{2i\omega t}\alpha_{3}|-,\tilde{+}\rangle)\label{eq22}
\end{align}
using Eq.(\ref{eq14}) and (\ref{eq21}), where the parameter $\omega $ denotes $\omega = \mu_{\text{B}}H/\hbar$.
This simple example shows that the entanglement between the original and tilde space decreases exponentially in the classical system shown in Eq.(\ref{eq20}).
This typical result corresponds to our previous study \cite{6}.

In some previous studies [7,20-22], the heat bath was introduced in a way different from the present study.
In those cases, the tilde Hamiltonian $\tilde{\mathcal{H}}_0$ is assumed as a heat bath directly \cite{7,21}, and the interaction between $\mathcal{H}_0$ and $\tilde{\mathcal{H}}_0$ affects as a coupling term $i\hbar S_j \tilde{S}_j$.
These assumptions may be plausible for large subsystems $\mathcal{H}_0$ including open quantum systems.
However, if we apply a small subsystem $\mathcal{H}_0$ as shown in this section, the tilde space $\tilde{\mathcal{H}}_0$ is too small to play a role of a heat bath.
In the present study, the heat bath is consistently introduced in the original space.
Then, reducing the thermal fluctuations both in the original and tilde spaces, the effective interaction $i\hbar \hat{\Lambda}(t)$ appears straightforwardly.
This is one of the most important points of the present perspective.

\section{Many-body systems affected by a heat bath}

In this section, we try to apply the perspectives shown in the previous section to many-body systems.
It is important to estimate how the effective thermal noises $\{R_j(t)\}$ affect the finite-size subsystem $\mathcal{H}_0$.
Here, we adopt an approach based on the cluster mean-field theory \cite{24}. 
At first, the relevant system is divided into two subsystems, namely the finite-size subsystem $\Omega$ and the infinite-size heat bath $\Omega'$.
These systems interact with each other through the interactions located in the boundary $\partial \Omega$.
Thus the Hamiltonian $\mathcal{H}$ is expressed as follows;
\begin{equation}
\mathcal{H}=\mathcal{H}_{\Omega}+\mathcal{H}_{\Omega'}+\mathcal{H}_{\partial \Omega}, \label{eq23}
\end{equation}
where the Hamiltonian  $\mathcal{H}_{\Omega}$ is the same as $\mathcal{H}_{0}$, namely $\mathcal{H}_{\Omega}=\mathcal{H}_0$.
On the scheme of the cluster mean-field theory \cite{24}, the fluctuations of parameters located in $\Omega'$ affect the parameters located in $\partial \Omega$ as effective fields.
When we assume the effective fields as random noises, we obtain the effective Hamiltonian
\begin{equation}
\mathcal{H}=\mathcal{H}_0-J\sum_{j\in \partial \Omega}R_j(t) S_j. \label{eq24}
\end{equation}
Here the spin parameters $\{S_j\}$ are located in the boundary region $\partial \Omega$.
The effective noises $\{R_j(t)\}$ satisfy the conditions
\begin{equation}
\langle R_j(t) \rangle_R =0 {\text{ and }} \langle R_i(t)R_j(t') \rangle_R = \epsilon_{ij} \delta(t-t') \delta_{ij},\label{eq25}
\end{equation}
where the parameters $\epsilon_{ij}$ and $\delta_{ij}$ denote the weights of covariances and Kronecker delta, respectively. 

Similar to section 2, we reduce the random noises included in the time development operator $\hat{U}(t)$.
We can derive the effective interaction $i\hbar \hat{\Lambda}(t)$ in the same way as shown in the previous section even for the present many body system because we can assume the noises $\{R_j(t)\}$ are independent of each other.
It is important to note that only noises are independent but the spins interact with each other.  
From Eq.(\ref{eq24}), the TFD Hamiltonian $\hat{\mathcal{H}}$ is derived as 
\begin{equation}
\hat{\mathcal{H}}=\hat{\mathcal{H}}_0-J\sum_{j\in \partial \Omega}R_j(t)(S_j-\tilde{S}_j)\equiv \hat{\mathcal{H}}_0+\hat{\mathcal{H}}'(t). \label{eq26}
\end{equation}
The moment $\langle\hat{\mathcal{H}}'(t)\hat{\mathcal{H}}'(t')\rangle_R $ is easily obtained as
\begin{align}
\langle\hat{\mathcal{H}}'(t)\hat{\mathcal{H}}'(t')\rangle_R &=J^2\sum_{j,k\in \partial \Omega}\langle R_j(t)R_k(t')\rangle_R (S_j-\tilde{S}_j)(S_k-\tilde{S}_k)\notag
\\
&=J^2\sum_{j,k\in \partial \Omega} \epsilon_{jk} (S_j-\tilde{S}_j)(S_k-\tilde{S}_k)\delta(t-t')\delta_{jk}\notag
\\
&=J^2\sum_{j\in \partial \Omega}\epsilon_{jj} (S_j-\tilde{S}_j)^2\delta(t-t').\label{eq27}
\end{align}
Inserting Eq.(\ref{eq26}) into Eq.(\ref{eq12}), we obtain the reduced time development operator with the decoupling approximation (\ref{eq13}) as
\begin{equation}
\langle \hat{U}(t)\rangle_{R}=e^{-i\hat{\mathcal{H}}_0t/\hbar}\exp\left[t\sum_{j\in \partial \Omega}\gamma_j ( S_j \tilde{S}_j-1) \right],\label{eq28}
\end{equation}
where the parameter $\gamma_j$ is defined as $\gamma_j=\epsilon_{jj}J^2/\hbar^2$.
Then the time development of the dissipative TFD state vector $|\Psi^{\text{diss}}(t) \rangle$ yields
\begin{equation}
\frac{\partial}{\partial t}|\Psi^{\text{diss}}(t) \rangle =\left(\frac{\partial}{\partial t}\langle \hat{U}(t)\rangle_{R}\right)|\Psi_0\rangle=\left(\frac{1}{i\hbar}\hat{\mathcal{H}}_0+\hat{\Lambda}(t) \right)|\Psi^{\text{diss}}(t) \rangle \label{eq29}
\end{equation}
and
\begin{equation}
\hat{\Lambda}(t) = \sum_{j\in \partial \Omega}\gamma_j \left( S_j(t)\tilde{S}_j(-t)-1 \right), \label{eq30}
\end{equation}
where $S_j(t)$ and $\tilde{S}_j(t)$ are defined as
\begin{equation}
S_j(t)\equiv e^{-i \mathcal{H}_0 t/\hbar}S_j e^{i \mathcal{H}_0 t/\hbar} {\text{ and }} \tilde{S}_j(t)\equiv e^{-i\tilde{\mathcal{H}}_0 t/\hbar}\tilde{S}_j e^{i\tilde{\mathcal{H}}_0 t/\hbar}. \label{eq31}
\end{equation}
This alternative relation is shown in Fig.2.

\begin{figure}[hbpt]
\begin{center}
\includegraphics[width=12cm]{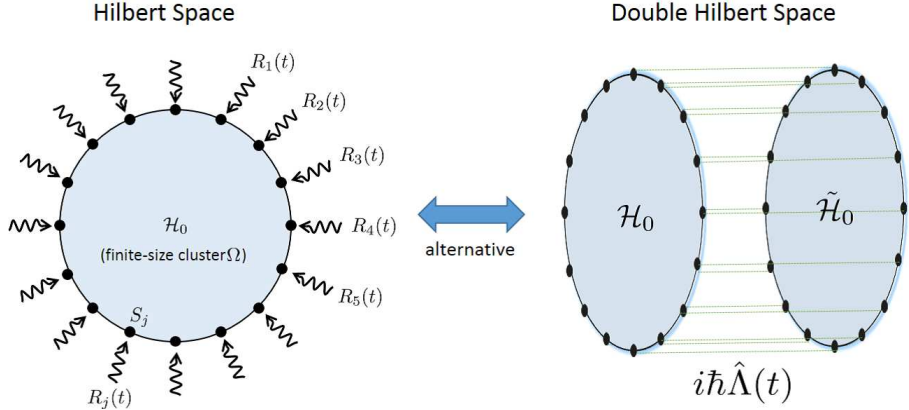}
\end{center}
\caption{Alternative relation of a many-body system}
\label{fig2}
\end{figure}

In the case where the picture of Fig.\ref{fig2} hold, the strength of the effective interaction $i\hbar \hat{\Lambda}(t)$ depends on the dimension $D$ as
\begin{equation}
|i\hbar \hat{\Lambda}(t)|\propto L^{D-1}\label{eq32}
\end{equation}
for the system size $L$ of the cluster $\Omega$, because Eq.(\ref{eq30}) shows that the effective interactions exist only in the boundary $\partial \Omega$.
On the other hand, the amount of the internal energy, $|\mathcal{H}_0|$, depends on the dimension as
\begin{equation}
|\mathcal{H}_0|\propto L^D, \label{eq33}
\end{equation}
because it is extensive.
Thus we can find that the thermal disturbance is relatively small.
Especially, for the thermodynamic limit $L\to \infty$, it becomes negligible.
In the present study, we have assumed that the thermal noises directly affect the spins on surface of the system, $\partial \Omega$, and they contribute gradually to the interior spins of the cluster $\Omega$ through the internal interactions.
Under these assumptions, we have shown by Eq.(\ref{eq30}) that the thermal noises can be reduced into the effective interaction between the surface spins of $\mathcal{H}_0$ and $\tilde{\mathcal{H}}_0$.
These perspectives of dissipative systems correspond to the fact that the interaction between the system and the heat bath is relatively very small but not zero, and this interaction leads the system to equilibrium states.

\section{Summary and discussions}

In the present study, we propose a new perspective on thermal dissipation as an effective interaction described by TFD theory.
Once thermal disturbances are reduced, there appear effective interactions between the original and tilde spaces.
Using the effective interactions, we show the correspondence between a dissipative model and a finite size system as shown in Fig.2.
By the way, as is well known \cite{25}, there appear thermal squeezed states on the dissipative TFD theory.
However, the present study does not consider the phase properties because it is not essential for the present discussions.
In fact, it is more important that the tilde space needs to be isomorphic to the original space, and then the same random noises appear in both of the original and tilde spaces, as shown in Eqs.(9) and (26).    
It enables us to obtain the effective interaction between the original and tilde spaces with reducing the thermal noises.
In the case of single free spin, the role of tilde space is to replace only mathematically the trace ${\text{Tr}}Q\rho(t)$ by the quantum average $\langle \Psi(t)|Q|\Psi(t) \rangle$.
Nevertheless, in a physical system with fluctuating forces, the tilde space is also integrated into the whole physical system $\hat{\mathcal{H}}$ with such effective interactions between the original and tilde spaces as it plays a role of thermal disturbance.
The present perspective may also be important to apply the TFD theory to thermal effects including entanglements and numerical studies.
The alternative system described in the double Hilbert space includes finite parameters while the heat bath includes infinite parameters.
Then the present perspective gives us a way to study the heat bath using finite size systems. 
It is useful for practical applications to study thermal effects, because in the present way with the effective interaction $i\hbar \hat{\Lambda}(t)$, only the same amount of computational effort is simply required as in finite-size systems.

\section{Acknowledgment}
One of the authors (Y.H.) is partially supported by JSPS Grant-in-Aid for Young Scientists (B) Grant Number 26800205.

\bibliographystyle{model1a-num-names}

\begin{thebibliography}{00}
\bibitem{1}
U. Fano, Rev. Mod. Phys. A42 (1957) 74.
\bibitem{2}
I. Prigogine, et al. , Chem. Scr. 4 (1973) 5.
\bibitem{3}
T. Takahashi and H. Umezawa, Collect. Phenom. 2 (1975) 55.
\bibitem{4}
M. Suzuki, "Quantum Fluctuation, Thermo Field Dynamics and Quantum Monte Carlo Methods" in the book "Progress in Quantum Field Theory", Chapter 10, Edited by H. Ezawa, and S. Kamefuchi, Elsevier, 1986.
\bibitem{5}
M. Suzuki, J. Stat. Phys. 42 (1986) 1047.
\bibitem{6}
Y. Hashizume and M. Suzuki, Physica A 392 (2013) 3518.
\bibitem{7}
A. E. Feiguin and S. R. White, Phys. Rev. B 72 (2005) 220401.
\bibitem{8}
G. Vidal, Phys. Rev. Lett. 99 (2007) 220405, ibid 101 (2008) 110501.
\bibitem{9}
G. Evenbly and G. Vidal, Phys. Rev. B 79, 144108 (2009); arXiv:1106.1082.
\bibitem{10}
H. Matsueda, M. Ishihara and Y. Hashizume, Phys. Rev. D 87 (2013) 066002.
\bibitem{11}
M. Nozaki, S. Ryu, and T. Takayanagi, J. High Energy Phys.@2012  (2012) 193.
\bibitem{12}
H. Matsueda, arXiv:1208.2872 [cond-mat.stat-mech].
\bibitem{13}
W. Israel, Phys. Lett. 57A (1976) 107.
\bibitem{14}
J. Maldacena, J. High Energy Phys. 4 (2003) 21.
\bibitem{15}
M. Suzuki, Int. J. Mod. Phys. B 5 (1991) 1821.
\bibitem{16}
H. Umezawa, Advanced Field Theory | Micro, Macro, and Thermal Physics, (AIP, New York,
1993).
\bibitem{17}
Y. Kuwahara, Y. Nakamura, Y. Yamanaka, Phys. Lett. A 377 (2013) 3102.
\bibitem{18}
M. Suzuki, J. Phys. Soc. Japan 54 (1985) 4483.
\bibitem{19}
M. Suzuki, in: L. Accardi, W. Freundberg, M. Ohya (Eds.), PQ-QP, QBIC, World Scientific, Singapore, 2008.
\bibitem{20}
T. Tominaga, M. Ban, T. Arimitsu, J. Pradko and H. Umezawa, Physica 149A (1988) 26
\bibitem{21}
H. Majima and A. Suzuki, J. Phys. Conf. Ser. 258 (2010) 012015.
\bibitem{22}
H. Majima and A. Suzuki, Ann. Phys. 326 (2011) 3000.
\bibitem{23}
J. Schwinger, J. Math. Phys. 2 (1961) 407.
\bibitem{24}
M. Suzuki, J. Phys. Soc. Japan 55 (1986) 4205.
\bibitem{25}
A. Kireev, A. Mann, M. Revzen and H. Umezawa, Phys. Lett. A 142 (1989) 215.
\end{thebibliography}

\end{document}